\documentclass[aip,twocolumn,superscriptaddress,pop,reprint]{revtex4-1}
\usepackage{amsmath}
\usepackage{amssymb}
\usepackage{mathtools}
\usepackage{natbib}
\usepackage{threeparttable}
\usepackage{booktabs}
\usepackage{bibentry}

\def\etal{{\it et al.}}
\def\eg{{\it e.g.,}}
\def\ie{{\rm i.e.,}}

\newcommand{\be}{\begin{equation}}
\newcommand{\ee}{\end{equation}}

\newcommand{\crash}{{\sc CRASH}}
\newcommand{\helios}{{\sc Helios}}
\newcommand{\propaceos}{{\sc Propaceos}}
\newcommand{\atbase}{{\sc ATBASE}}

\newcommand{\TR}{{T$_{\rm R}$}}

\newcommand{\SN}{{S$_{n}$-nLTE}}
\newcommand{\NI}{N$\;${\small{\rm I}}\relax}
\newcommand{\NII}{N$\;${\small{\rm II}}\relax}
\newcommand{\NIII}{N$\;${\small{\rm III}}\relax}
\newcommand{\NIV}{N$\;${\small{\rm IV}}\relax}
\newcommand{\NV}{N$\;${\small{\rm V}}\relax}
\newcommand{\NVI}{N$\;${\small{\rm VI}}\relax}
\newcommand{\NVII}{N$\;${\small{\rm VII}}\relax}
\newcommand{\NVIII}{N$\;${\small{\rm VIII}}\relax}

\newcommand{\D}{\textrm{d}}

\begin{document}
\title{Atomic Modeling of Photoionization Fronts in Nitrogen Gas}
\author{William J. Gray}
\affiliation{CLASP, College of Engineering, University of Michigan, 2455 Hayward St., Ann Arbor, Michigan 48109, USA}
\author{P. A. Keiter}
\affiliation{CLASP, College of Engineering, University of Michigan, 2455 Hayward St., Ann Arbor, Michigan 48109, USA}
\affiliation{Los Alamos National Lab, Los Alamos, New Mexico, 87544}
\author{H. Lefevre}
\affiliation{CLASP, College of Engineering, University of Michigan, 2455 Hayward St., Ann Arbor, Michigan 48109, USA}
\author{C. R. Patterson}
\affiliation{CLASP, College of Engineering, University of Michigan, 2455 Hayward St., Ann Arbor, Michigan 48109, USA}
\author{J. S. Davis}
\affiliation{CLASP, College of Engineering, University of Michigan, 2455 Hayward St., Ann Arbor, Michigan 48109, USA}
\author{K. G. Powell}
\affiliation{Department of Aerospace Engineering, University of Michigan, Ann Arbor, Michigan, 48109, USA}
\author{C. C. Kuranz}
\affiliation{CLASP, College of Engineering, University of Michigan, 2455 Hayward St., Ann Arbor, Michigan 48109, USA}
\author{R. P. Drake}
\affiliation{CLASP, College of Engineering, University of Michigan, 2455 Hayward St., Ann Arbor, Michigan 48109, USA}

\begin{abstract}
	Photoionization fronts play a dominant role in many astrophysical environments, but remain difficult to achieve in a laboratory experiment. Recent papers have suggested that experiments using a nitrogen medium held at ten atmospheres of pressure that is irradiated by a source with a radiation temperature of T$_{\rm R}\sim$ 100 eV can produce viable photoionization fronts. We present a suite of one-dimensional numerical simulations using the \helios\ multi-material radiation hydrodynamics code that models these conditions and the formation of a photoionization front. We study the effects of varying the atomic kinetics and radiative transfer model on the hydrodynamics and ionization state of the nitrogen gas, finding that more sophisticated physics, in particular a multi-angle long characteristic radiative transfer model and a collisional-radiative atomics model, dramatically changes the atomic kinetic evolution of the gas. A photoionization front is identified by computing the ratios between the photoionization rate, the electron impact ionization rate, and the total recombination rate. We find that due to the increased electron temperatures found using more advanced physics that photoionization fronts are likely to form in our nominal model. We report results of several parameter studies. In one of these, the nitrogen pressure is fixed at ten atmospheres and varies the source radiation temperature while another fixes the temperature at 100 eV and varied the nitrogen pressure. Lower nitrogen pressures increase the likelihood of generating a photoionization front while varying the peak source temperature has little effect.  
\end{abstract}
\maketitle

\section{Introduction}

Photoionization fronts play an essential role in the evolution of the Universe at all scales.\citep{Robertson2010} The Universe was plunged into the so-called Dark Ages, a time after hydrogen and helium recombined after the Big Bang and no sources of emission existed. It is during this time that the first structures began to form.\citep{Press1974} Comprised of mostly dark matter with virial masses of M$_{vir}\sim$10$^{6}$M$_{\odot}$\citep{Greif2015}, not only are these minihalos the fundamental building blocks of larger structures, they are the birthplace of the first stars. \citep{Bromm2004,Bromm2013,Clark2011,Glover2005,Glover2013,Bromm2017} These Population III stars have masses of M$_{*}\sim$100 M$_{\odot}$ and are more compact, have higher surface temperatures, and produce many more UV photons compared to present-day stars.\citep{Tumlinson2000} These first stars also provided the first metals that enriched the intergalactic medium.\citep[\eg][]{Yoshida2004,Jaacks2018}

The first galaxies formed through the hierarchical merging of these minihalos. Once the mass of the dark matter halo reaches $\sim$10$^{8}$ M$_{\odot}$, the virial temperature of the minihalo is greater than the atomic hydrogen cooling limit and a population of stars can form.\citep{Oh2002,Wise2007,Wise2008,Greif2008,Greif2010,Bromm2011,Clark2011,Greif2015}  These stars were then essential in heating and ionizing the surrounding gas, forming photoionization fronts that created additional structure within these ``minihalos''. The end of the Dark Ages then came about as these stars, in collaboration with black holes, proceeded to reionize the universe. 

Photoionization fronts also play an important role in the present day universe. Massive stars, M$_{*}>$20 M$_{\odot}$(\eg\ O-type stars), have high surface temperatures and produce a large number of ionizing photons. H II regions are created as these photons ionize the gas surrounding these stars.\citep[\eg][]{Franco2000,Williams2000,Mackey2016,Gvaramadze2017} These photoionization fronts have important effects on the surrounding gas, from destroying giant molecular clouds, affecting future star formation, and even limiting the final stellar mass of a high mass protostellar object.\citep{McKee2007} Photoionization fronts are also important in the transition between ``pre-planetary'' to ``planetary'' nebula during the late-stage evolution of intermediate mass stars ($\sim$0.8-8 M$_{\odot}$).\citep[\eg][]{Tafoya2013,Sabin2014,Gledhill2015,Planck2015}

Until recently there have only been a handful of laboratory experiments potentially relevant to photoionization fronts. Most have made use of carbon foams or low-density plastics illuminated by x-ray sources. In the experiments by Willi and collaborators \citep{Afsharrad1994,Hoarty1999a,Hoarty1999b,Willi2000} used low-density triacrylate foams irradiated by a soft x-ray source generated by laser irradiation of a gold `burn-through' foil. The authors used x-ray spectroscopy and radiography to infer density and temperature profiles through the foam and were suggestive of a supersonic ionization front. Similarly, Zhang \etal\ \cite{Zhang2010} used a low-density plastic (C$_{8}$H$_{8}$) foam with x-rays generated by irradiating the interior of a gold cylinder. Using time-resolved x-ray radiography the authors measured ionization and shock positions within their foams. These were then compared to numerical simulations and an ionization front was found to precede the shock front at early time. Drake \etal \cite{Drake2016}, however, showed that these experiments did not produce photoionization fronts, but rather heat fronts where the energy is transported through electron heat conduction.

Drake \etal \cite{Drake2016} outlined the theoretical requirements for an authentic laboratory photoionization front experiment. Although a variety of materials were studied, nitrogen gas held at high pressure was found to be an appropriate medium for a photoionization front experiment. Their study suggested from analytical calculations that a radiation source of finite diameter and a temperature of T$_{\rm R}$=100 eV irradiating nitrogen gas held at ten atmospheres is sufficient to generate a photoionization front.
The high temperature is necessary to produce the large photon flux needed in a laboratory setting.
A consequence is that the photon spectrum required in the laboratory is much harder than that of the ionizing stellar radiation in nature.
Gray \etal \cite{Gray2018} followed this up by using two-dimensional \crash\ \citep{vanderHolst2011} simulations to study the proposed photoionization front experiment. These authors studied two experimental setups, one where the x-ray source is generated from a laser heated foil and a pulse powered source where the source is generated from the implosion of current carrying wires. They found that a photoionization front is formed in the laser heated foil for moderate nitrogen pressures (five to ten atm) and high radiation source temperatures T$_{\rm R}>$90 eV. In the pulsed power simulations, the radiation source temperature is fixed at T$_{\rm R}=$90 eV and lower nitrogen pressures are favored, two to five atmospheres. However, these simulations are limited to flux limited diffusion radiation transport and a local thermodynamic equilibrium atomic kinetics model to compute the ionization states.

As mentioned in Drake \etal \cite{Drake2016} the primary photoabsorption mechanism is photoelectric, that is, the freed electron has an energy equal to the photon energy minus the ionization energy. In astrophysical systems, stars providing the ionizing photons have temperatures of a few eV and only the photons in the high-energy tail drive the photoionization front. Therefore, the freed electrons have an energy of only a few eV. In contrast, laboratory sources have temperatures much higher than the characteristic ionization energies, which translate to much higher electron energies. However, as stressed in Drake \etal \cite{Drake2016}, even if these laboratory experiments fail to precisely match the astrophysical conditions, a structured experiment of a photoionization front is a meaningful first step. 

In the present paper, we aim to expand on the simulations presented in Gray \etal \cite{Gray2018} and study the effects of more sophisticated radiation transport and atomic kinetics models. To that end, we present here a suite of simulations using the one-dimensional Lagrangian code \helios.\citep{MacFarlane2006} We compare the effect of changing the radiative transfer model and the atomic kinetics model. We also perform a parameter study where the source radiation temperature is varied for a fixed nitrogen pressure and one where the nitrogen pressure is varied for a fixed source radiation temperature. 

The structure of the paper is as follows. In \S 2 we present the simulation framework and initial conditions. In \S 3 we present the results of varying the atomic kinetics, particle resolution, and radiative transfer model. \S 4 presents the results of our parameter study, in particular whether or not a photoionization front is generated. Summary and conclusions are given in \S 5. 

\section{Model Framework and Initial Conditions}
\begin{figure}
\begin{center}
\includegraphics[trim=0.0mm 0.0mm 0.0mm 0.0mm, clip, width=0.85\columnwidth]{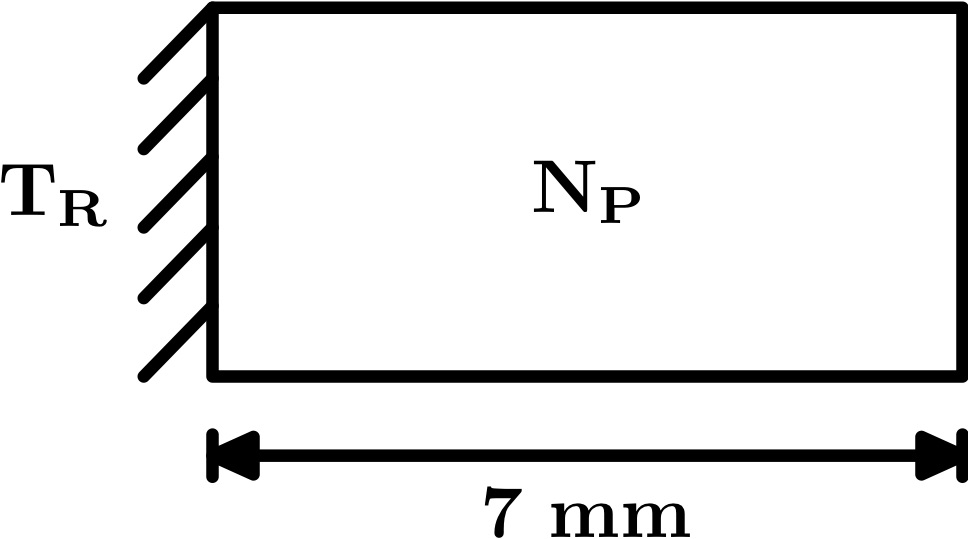}
\caption{Schematic view of the experiment setup. The energy input on the left boundary is parameterized by \TR\ while the gas cell is parameterized by the nitrogen pressure, N$_{\rm P}$. The length of the gas cell is fixed at 7 mm. }
\label{fig:schematic}
\end{center}
\end{figure}

\begin{table*}
\begin{centering}
\resizebox{0.95\textwidth}{!}{%
\begin{threeparttable}
	\caption{Simulation Summary}
	\label{tab:runsummary}
	\begin{tabular}{|l||cc|cc|cc|c|}
	\hline
	Name                & N$_{P}$ (atm)    & Peak T$_{R}$ (eV) & S$_{n}$    & FLD          & nLTE       & LTE          & Resolution \\
	\hline
	\hline
	Sn-nLTE (nominal)   & 10               & 100               & \checkmark &  -           & \checkmark & -            & 200        \\
	FLD-LTE             & 10               & 100               & -          &  \checkmark  & -          & \checkmark   & 200        \\
	Sn-LTE              & 10               & 100               & \checkmark &  -           & -          & \checkmark   & 200        \\
	\hline
	Sn-nLTE-05x         & 10               & 100               & \checkmark &  -           & \checkmark & -            & 100        \\
	Sn-nLTE-2x          & 10               & 100               & \checkmark &  -           & \checkmark & -            & 300        \\
	\hline
	Sn-nLTE-80eV        & 10               & 80                & \checkmark &  -           & \checkmark & -            & 200        \\ 
	Sn-nLTE-90eV        & 10               & 90                & \checkmark &  -           & \checkmark & -            & 200        \\ 
	Sn-nLTE-120eV       & 10               & 120               & \checkmark &  -           & \checkmark & -            & 200        \\ 
	Sn-nLTE-140eV       & 10               & 140               & \checkmark &  -           & \checkmark & -            & 200        \\
	\hline
	Sn-nLTE-2.5atm      & 2.5              & 100               & \checkmark &  -           & \checkmark & -            & 200        \\ 
	Sn-nLTE-5atm        & 5                & 100               & \checkmark &  -           & \checkmark & -            & 200        \\ 
	Sn-nLTE-20atm       & 20               & 100               & \checkmark &  -           & \checkmark & -            & 200        \\
	Sn-nLTE-40atm       & 40               & 100               & \checkmark &  -           & \checkmark & -            & 200        \\
\hline
\end{tabular}
\begin{tablenotes}
\item	{\bf Notes:} Synopsis of \helios\ models. The first column gives the simulation name, the second column gives the initial uniform nitrogen pressure, and the third column gives the peak radiation temperature used as the boundary condition. The next four columns give the choice of physics packages where a checkmark indicates the implemented option. The number of Lagrangian particles used is given in the last column.
\end{tablenotes}
\end{threeparttable}
}
\end{centering}
\end{table*}

All of the simulations presented here are performed using \helios\ (v7.4.0), a 1-D Lagrangian multi-material radiation-hydrodynamics code { developed by Prism Computational Sciences Inc}.\citep{MacFarlane2006} A variety of atomic and radiation physics options are available in \helios. Table~\ref{tab:runsummary} gives a summary of the simulations run and the various physics options used. In particular, we vary the radiative transfer method between a flux limited diffusion (FLD) model and a multi-angle long characteristic transport model (S$_n$). In both cases multi-frequency radiative transfer is used with thirty radiation groups. { These are set up} using the default options as implemented by \helios; which puts 85\% of the energy groups between 0.1 eV and 3 keV and the remaining 15\% between 3 keV and 1 MeV. 

Both flux limited diffusion and multi-angle long characteristics model have their advantages and disadvantages. The diffusion approximation is well suited to optically thick media and is computationally efficient. FLD also makes use of a flux limiter that ensures the radiation does not propagate faster than the speed of light. These flux limiters also allow for an approximate way of moving between optically thin and optically thick regimes.\citep[\eg][]{Levermore1981,vanderHolst2011}

The long characteristics model does not do a diffusion calculation, but instead directly solves the time independent radiative transfer equation based on the method introduced in Olson \etal\cite{Olson1987} As discussed in MacFarlane \etal \cite{MacFarlane2006}, this solves these equations on a discretized grid of optical depths. Thus, this method properly captures the radiative transfer in optically thick, thin, and intermediate regimes. However, this complexity comes at a higher computational cost.

Two options are used for the atomic physics, local (LTE) and non-local (nLTE) thermodynamic equilibrium. In the case of LTE, the opacities and equation of state properties are provided by lookup tables generated by \propaceos\ (see Appendix A of MacFarlane \etal \cite{MacFarlane2006}). In the non-LTE case, an in-line collisional-radiative model is employed. Here, the atomic level populations are computed using a coupled set of atomic rate equations. Atomic cross section data is obtained from the \atbase\ \citep{Wang1991} suite of atomic codes. The total number of atomic levels computed by \helios\ can be varied, here we employ standard ``CR'' atomic model provided by \atbase. {\ Golovkin \etal\ \citep{Golovkin2003} showed that this model converges to LTE conditions when it should.\footnote{Golovkin, Igor, Internal Prism Scientific Report, Private Communication.} } In addition, a frequency grid is created that resolves the bound-bound and bound-free transitions and the transport equation is evaluated for each frequency point. We note that the choice of radiative transfer model and atomic model are independent. That is, it is possible to run \helios\ with S$_{n}$ radiative transfer and LTE atomic kinetics. The combination of these options is explored below.

Figure~\ref{fig:schematic} shows a schematic view of our initial conditions. Several coordinate systems are available in \helios, we choose to run each simulation in a planar coordinate system as it best matches the experimental design. A radiation temperature boundary condition is imposed on the left boundary. The initially zero radiation temperature is linearly ramped up to a peak value over the first nanosecond and then held fixed at the peak value, \TR, for the remaining simulation runtime. This profile is chosen to model the results { reported} in Davis \etal\cite{Davis2016} These authors studied the back side emission from a laser illuminated 0.5 {$\mu$m} gold foil, finding a rise time of about one nanosecond and a peak { radiation brightness} temperature of T$_{\rm R}\sim$100 eV. The gas cell is filled with nitrogen gas fixed at a target pressure, $N_P$. Nominally, 200 Lagrangian points are used to resolve the gas cell with default gridding provided by \helios. In addition, fixed hydrodynamic boundary conditions are employed for the first and last grid points, that is, they are held fixed at their initial positions. Following the setup used in Gray \etal \cite{Gray2018} the nitrogen gas cell is 7 mm in length.

The four sets of simulations performed are described in Table~\ref{tab:runsummary}. The first set aims at determining the importance between the atomic kinetics model (LTE versus nLTE) and radiation transfer method (FLD versus S$_{n}$). A resolution study is performed with a second set of simulations. Finally, a parameter study is performed with the last two sets of simulations. The first studies the impact of changing the peak radiation boundary condition temperature, from between 80 to 140 eV, while keeping the nitrogen pressure constant at ten atmospheres. The final set of simulations varies the nitrogen pressure from 2.5 to 40 atmospheres of nitrogen and using a peak radiation temperature of 100 eV.

\section{Results}

\subsection{Nominal Model Results}
\label{sec:nominalmodel}
\begin{figure*}
\begin{center}
\includegraphics[trim=0.0mm 0.0mm 0.0mm 0.0mm, clip, width=0.85\textwidth]{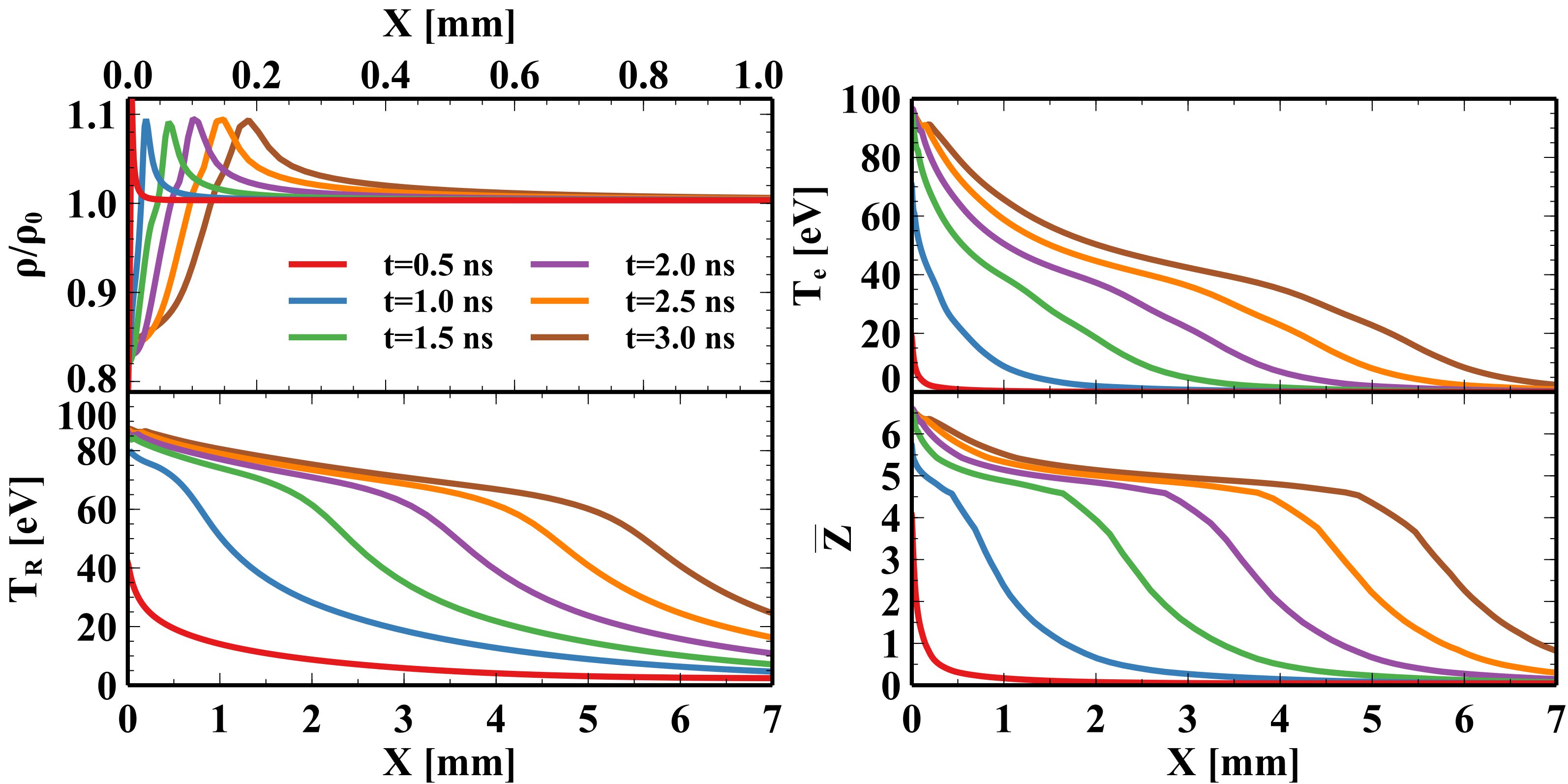}
\caption{Atomic and hydrodynamic evolution of the nominal model. {\it Top left}: shows the normalized mass density, {\it top right}: shows the electron temperature in eV, {\it bottom left}: shows the radiation temperature in eV, and {\it bottom right}: shows the average atomic ionization state. Note that the mass density uses a top axis which extends to 1 {mm} while all other variables are plotted using an axis that extends to 7 {mm}.  The mass is normalized by the initial ambient nitrogen gas density. }
\label{fig:fiducial}
\end{center}
\end{figure*}

We define our nominal model as a model with ten atmospheres of nitrogen, a peak radiation temperature of 100 eV, 200 Lagrangian points, S$_{n}$ radiative transfer, and non-LTE atomic kinetic physics. Each simulation is run for a total of 3 nanoseconds, which allows for the radiative temperature on the boundary to reach its peak value and emit for 2 nanoseconds.

Figure~\ref{fig:fiducial} shows the evolution of hydrodynamic and atomic physics variables as a function of time. The top left panel shows $\rho/\rho_{0}$ where $\rho_{0}$ is the initial ambient nitrogen density. Importantly, the peak density is only $\sim$10\% greater than the initial density, which suggests that a strong hydrodynamic shock does not form in the nitrogen gas. This matches the both the analytic and numerical results found in Drake \etal \cite{Drake2016} and Gray \etal\cite{Gray2018} In contrast to the mass density, both the radiation and electron temperature evolve to very high values at relatively early times. However, the radiation temperature evolves faster and produces higher peak temperatures compared to the electron temperature. Finally, the gas reaches a nearly uniform ionization state of \NVI\ and radiation temperature, \TR\, where \TR\ is the temperature of a Planckian spectrum whose energy density equals the energy density of the radiation as found in the simulation (\TR$\sim$80 eV).

\subsection{Atomic Model and Radiation Model Effects}
\label{sec:atomicmodels}
\begin{figure*}
\begin{center}
\includegraphics[trim=0.0mm 0.0mm 0.0mm 0.0mm, clip, width=0.85\textwidth]{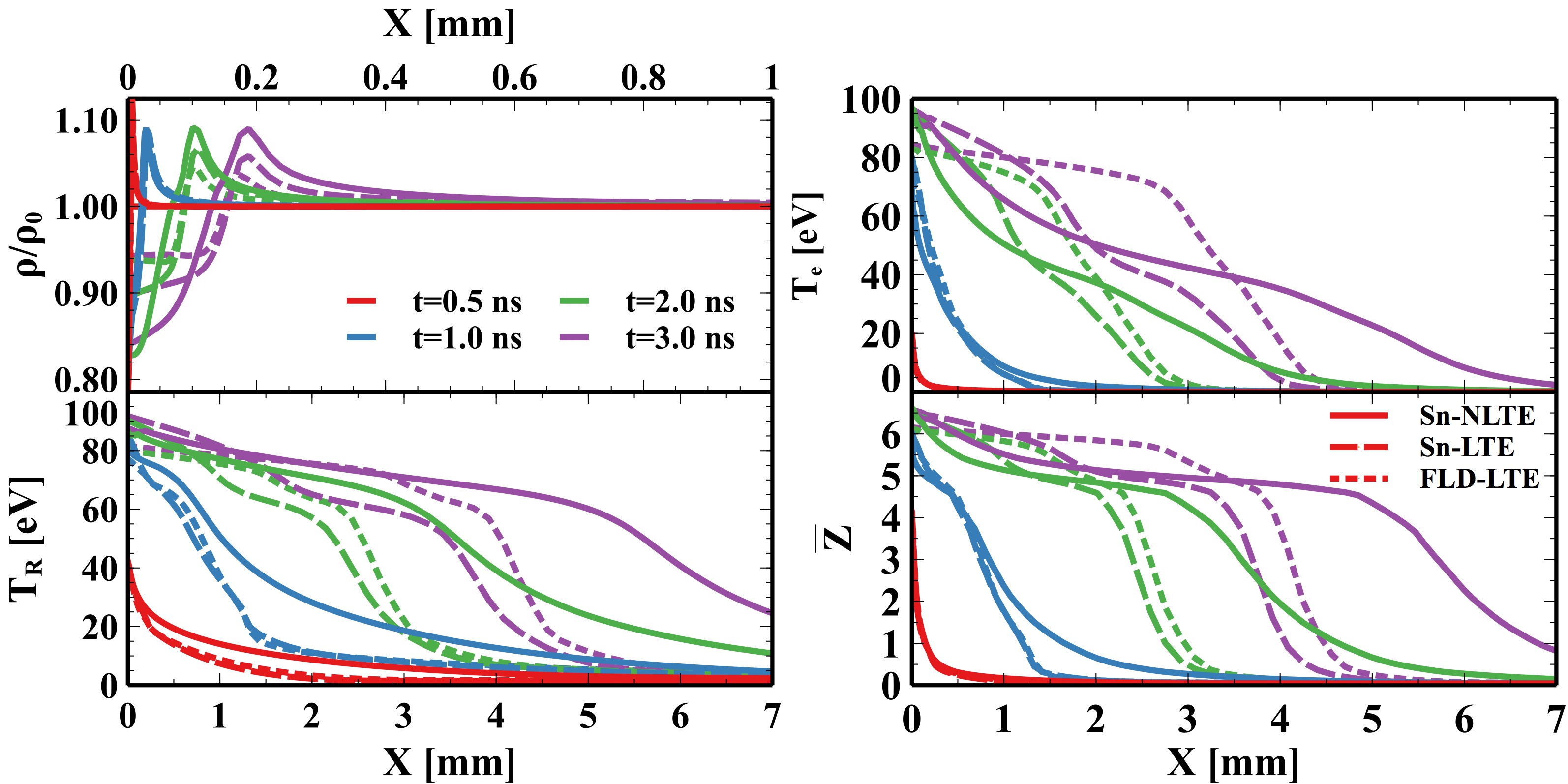}
\caption{ Comparison of simulations that vary the atomic physics and radiation transport. {\it Top Left}: Ratio of the mass density to the ambient density, {\it Top Right}: electron temperature in units of eV, {\it Bottom Left}: radiation temperature in units of eV, and {\it Bottom Right}: the average ionization state. Note that the mass density uses a top axis which extends to 1 {mm} while all other variables are plotted using an axis that extends to 7 {mm}. The legend gives the models considered and are described in Table~\ref{tab:runsummary}. }
\label{fig:atomicmodel}
\end{center}
\end{figure*}

\begin{figure}
\begin{center}
\includegraphics[trim=0.0mm 0.0mm 0.0mm 0.0mm, clip, width=0.45\textwidth]{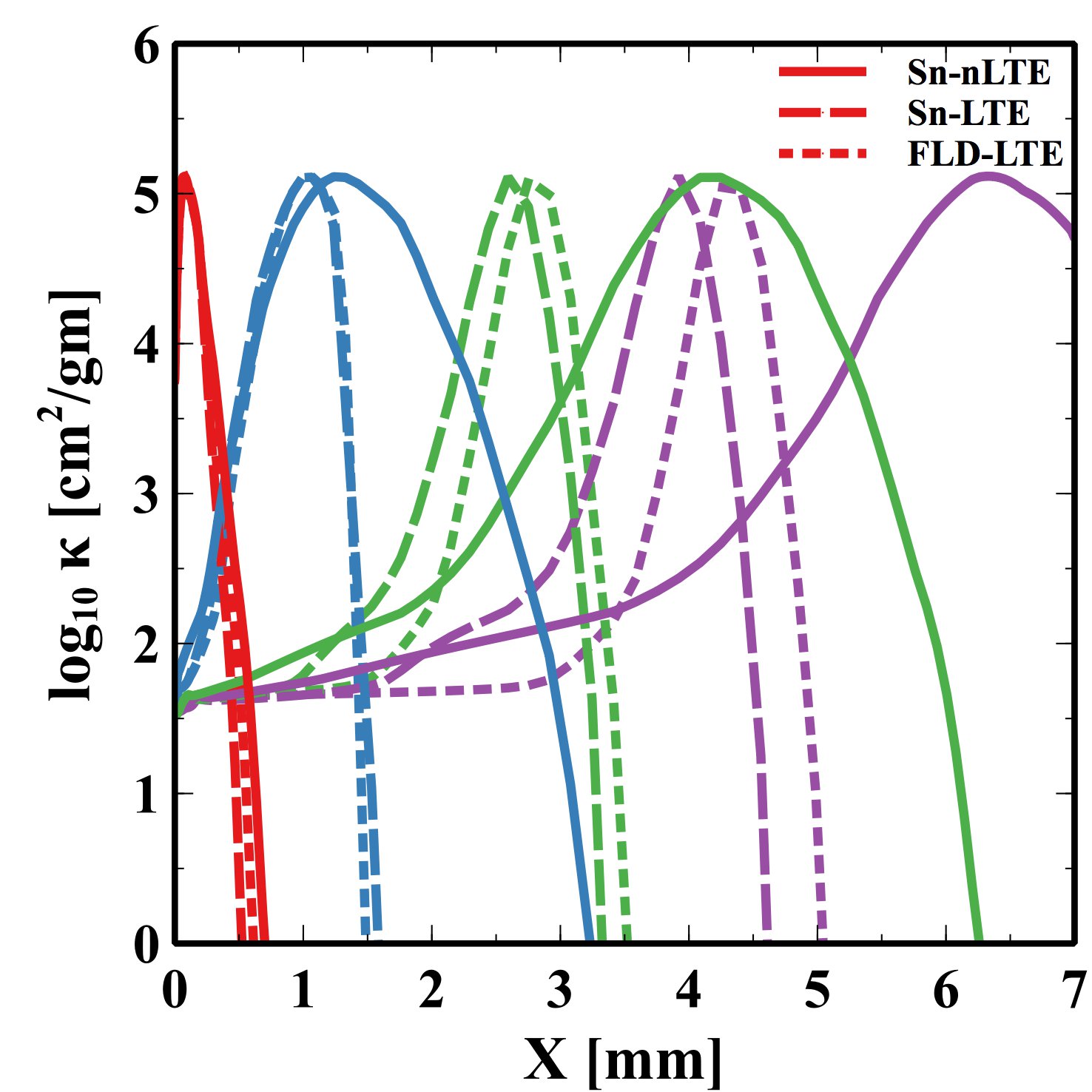}
\caption{ Comparison of the Rosseland mean opacities. The line style represents each model and is given by the legend. $t$=0.5 ns is shown by the red line, $t$=1.0 ns is shown by the blue, $t$=2.0 ns is shown by the green, and $t$=3.0 ns is shown by the purple line. The $x$-axis gives the position along the gas cell while the $y$-axis gives the logarithm of the opacity.  }
\label{fig:opacity}
\end{center}
\end{figure}

In this section we study the effect that varying the atomic model has on the hydrodynamics and atomic properties. Two options are used for radiation transport, flux limited diffusion, denoted as FLD, with the Larsen flux limiter \citep{Levermore1981} and a multi-angle long characteristic model based on Olson \etal \cite{Olson1987}, denoted as S$_{n}$. Similarly, two options are used for the atomic physics;  local (LTE) and non-local (nLTE) thermodynamic equilibrium. LTE ionization state level populations can be precomputed and are provided via table lookup. nLTE ionization state values are computed using {the} in-line collisional radiative {model described above}. The choice of radiative transfer model (FLD or S$_n$) and atomic kinetics model (LTE or nLTE) therefore describes each simulation. {We note that only the radiative transfer model and the atomic kinetics model are varied here. Each model uses the same initial nitrogen pressure and radiation temperature boundary condition (\ie\ same peak radiation temperature and same temporal profile). }

Figure~\ref{fig:atomicmodel} shows the results for this set of simulations. In order of increasing complexity they are FLD-LTE, S$_{n}$-LTE, and S$_{n}$-nLTE. Each simulation is compared at four times, at $t=0.5$ ns when the source is half way to its peak value, at $t=1.0$ ns when it has reached its peak, and at $t=2.0$ ns and $t=3.0$ ns when the source is emitting at full strength. We note that while the source temperature reaches its peak value, the actual brightness temperature at the opening to the gas cell is lower than that of the source. This is due to radiative absorption and the multi group nature of the radiative transfer.

The top left panel of Figure~\ref{fig:atomicmodel} shows the ratio of the mass density to the ambient density. As expected for a photoionization front \citep{Drake2016,Gray2018} there is little to no density evolution over the length of the simulation. These density profiles, including the density peaks, are nearly identical for these models, amounting to a difference of only a few percent. Neither the atomic physics or radiation transport method has much influence on the hydrodynamics for these simulations.

The top right panel shows the electron temperature in units of eV for each simulation. At early times, $t<$1 ns, all three simulations match well. However, by 2 ns, a clear separation between the different models is found. { First consider the impact of the} radiation transport method. The S$_{n}$ model produces a much shallower electron temperature profile when compared to the FLD model. Importantly, the FLD model produces a much hotter environment compared to the long characteristic simulation. { This difference is in the direction one would expect. The radiation mean free path, as a fraction of the temperature scale length near the front, is not small enough to make FLD strictly valid. In this regime, it would be expected to create an excessive amount of radiation transport, and this is what has occurred.}

{ Now consider the effect of the choice of atomic model while using S$_{n}$ radiation transport. This} produces a drastic difference in the electron temperature. Although S$_{n}$-nLTE shares the same shallow electron temperature profile as S$_{n}$-LTE, it produces appreciably higher temperatures further into the nitrogen gas cell, $\sim$ 5 mm. In fact, in regions where the electron temperature is negligible in S$_{n}$-LTE, the electron temperature can be as high as 40 eV in S$_{n}$-nLTE. { This is a natural result of the difference in atomic kinetics. When photoionization is larger than it would be in an LTE plasma with the given electron temperature, the consequence is that the plasma becomes more ionized than it would be in LTE at that electron temperature. This has the effect of reducing the opacity at the energies relevant to thermal ionization, which in turn allows the radiation to penetrate farther. Both Figure 4 below and our later analysis of ionization rates below supports this interpretation. }

The bottom left panel shows the radiation temperature in units of eV for each simulation. { The variations seen here are consistent with our interpretation of the variations in electron temperature. When using the LTE atomic kinetics model, the results do not vary strongly with the radiation transport method. } Both FLD-LTE and S$_{n}$-LTE largely track each other with FLD-LTE permeating slightly further into the nitrogen gas when compared at the same times. Similarly the peak radiation temperatures are very similar. The effect the collisional-radiative modeling on the radiation temperature is very similar to that found in the electron temperature. At nearly all times, S$_{n}$-nLTE predicts that the radiation has penetrated all the way through the nitrogen gas, producing temperatures of up to $\sim$30 eV at the end of the gas cell. { Just as in the case of the electron temperature, this is a natural result of the affect of strong photoionization on the opacity. }

Finally, the bottom right compares the average charge for each simulation. {The ionization state for all three models is similar across most of the heated region. This suggests that the equation of state of the ionized material is not the main origin of differences between the models. Instead, we attribute the differences in ionization profiles to the differences in opacity whose origin is described above.} As found for the other variables discussed above, S$_{n}$-nLTE shows the largest difference compared to the other the methods.  In particular, ionized gas is found up to 2 mm farther down the gas cell in S$_{n}$-nLTE compared to either S$_{n}$-LTE or FLD-LTE.  All three simulations produce a quick rise in ionization state and a plateau near the \NVI\ state. However, in S$_{n}$-nLTE this plateau is much more pronounced, stretching from 1 to 5.5 mm at $t=$3 ns.

These models highlight the dramatic differences seen when using a non-equilibrium atomic kinetics model. In particular, the radiation is able to penetrate much farther down the gas cell when compared to the LTE model. { The Helios code provides the Rosseland mean opacity as an available output parameter. Figure 4 shows profiles from the various models at various times. It confirms our expectations from the discussion above. At later times, at the locations where opacity in the the LTE models is near its peak, the value from the nLTE model is smaller by up to a few orders of magnitude. Our interpretation is that this is a result of the increase of the ionization state by photoionization, beyond the level that corresponds to LTE behavior at the local electron temperature. This in turn enables deeper penetration of the radiation energy flux. }

\subsection{Resolution Study}
\label{sec:resolution}
{
To ensure that our results are not resolution dependent, we have run a set of models that vary the number of Lagrangian particles. Two additional models are run, one with 100 particles and one with 300 particles using the same parameters as Sn-nLTE. We find negligible differences runs with 300 and 200 particles and small differences between the 200 and 100 particle runs.The nominal value of 200 particles is therefore sufficient to resolve the hydrodynamics and important physics considered here.
}

\section{Photoionization Front}
\label{sec:photo}

\begin{figure*}
\begin{center}
\includegraphics[trim=0.0mm 0.0mm 0.0mm 0.0mm, clip, width=0.85\textwidth]{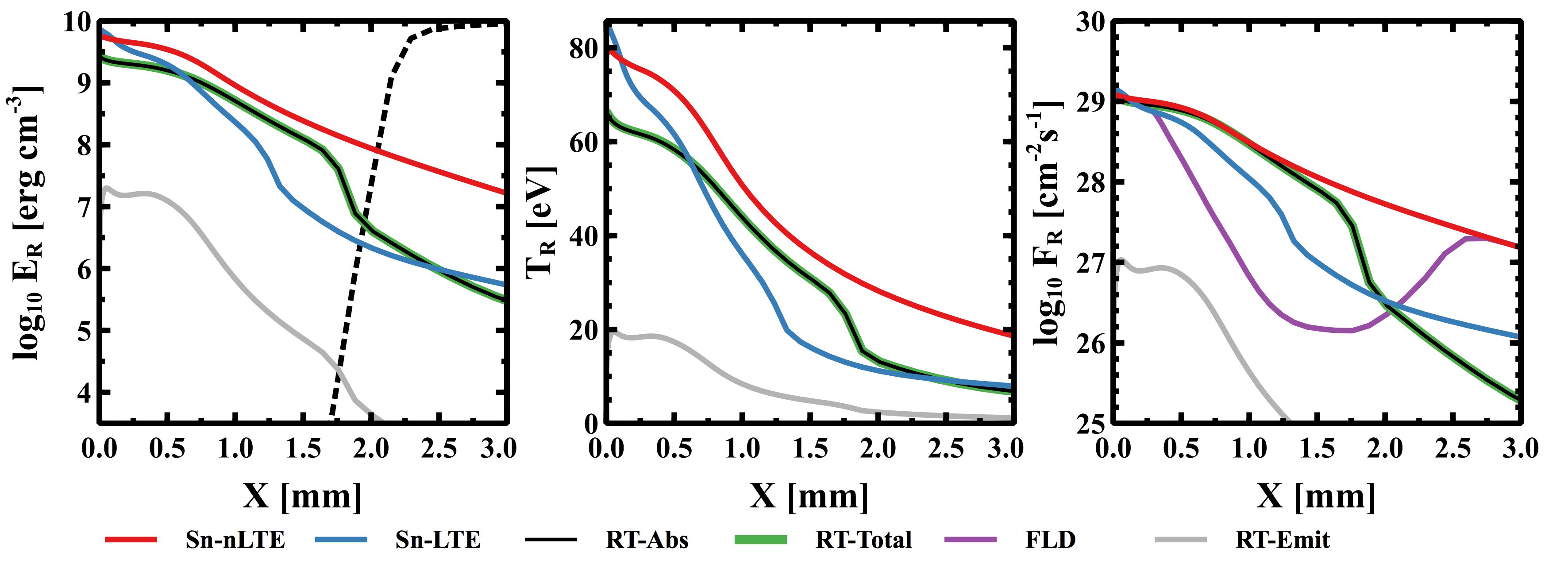}
\caption{ Ray trace results showing the {\it left panel: } radiation energy density, {\it middle panel: } radiation temperature in units of eV, and the {\it right panel: } photon flux. \helios\ results are shown by the red and blue lines for \SN\ and S$_n$-LTE models respectively. The black line shows the absorption only results while the light gray line shows the thermal emission of the nitrogen gas. The sum of these two components is shown by the green line. Note that the absorption component dominates over the thermal emission and causes the total emission line to lie on top of the absorption line. The dashed black line shows the neutral nitrogen ionization fraction where everything to the left of this line is ionized. A photoionization front may exist for neutral nitrogen at the region where this line slopes steeply upward and to the left. { The purple curve for photon flux is explained in the text.} }
\label{fig:raytrace}
\end{center}
\end{figure*}

\begin{figure}
\begin{center}
\includegraphics[trim=0.0mm 0.0mm 0.0mm 0.0mm, clip, width=0.85\columnwidth]{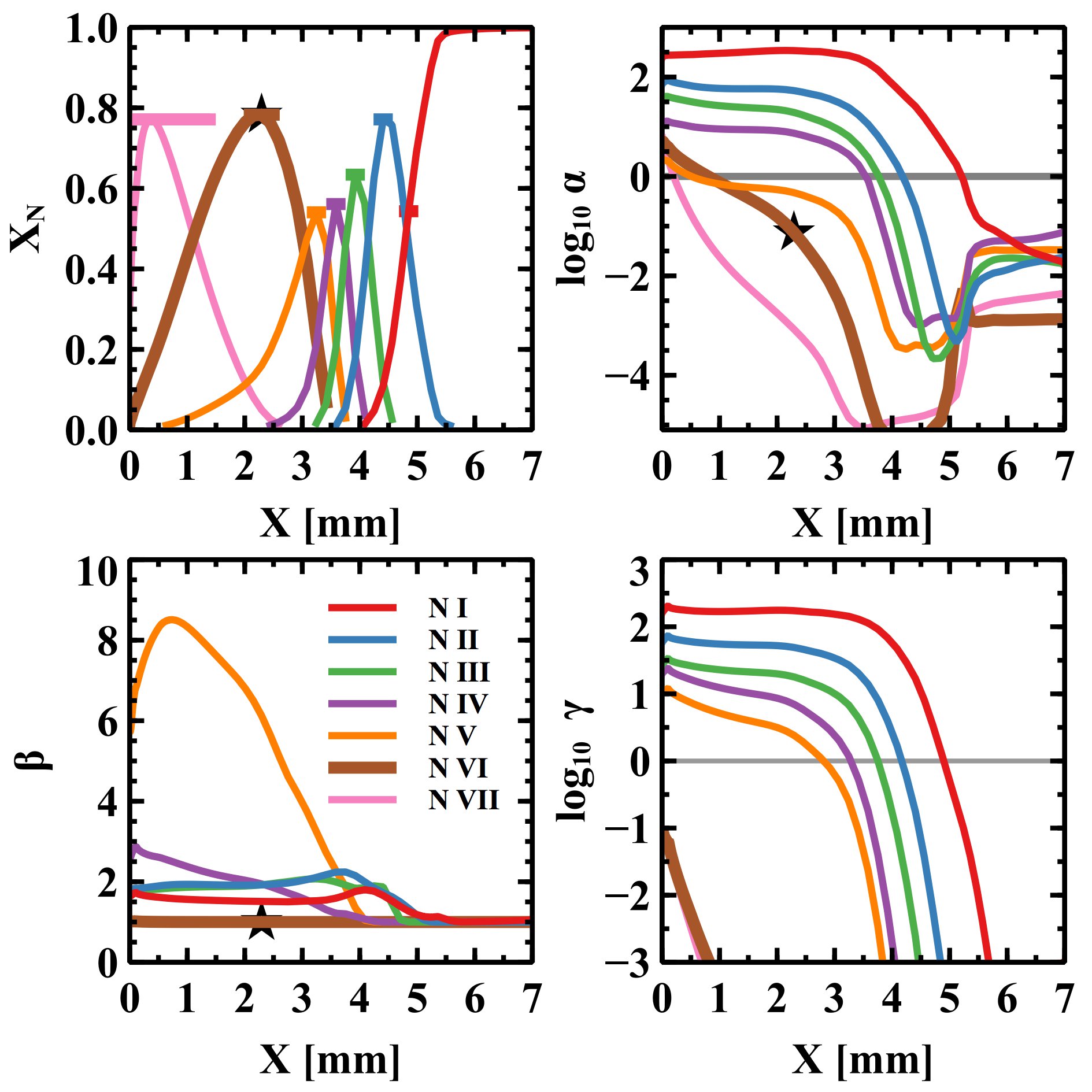}
\caption{ {\it Top left: } Ionization states of nitrogen, {\it Top right: } $\alpha$, {\it Bottom left: } $\beta$, and {\it Bottom right: } $\gamma$ values as a function of position for our nominal model at 2 ns. The black star represents the position where \NVI\ is maximum. Line color is consistent in each panel and represents a given ionization state. The horizontal bars above the maximum of each ionization state is represents the mean free path for that ionization state.}
\label{fig:alphabetapos}
\end{center}
\end{figure}

\begin{figure*}
\begin{center}
\includegraphics[trim=0.0mm 0.0mm 0.0mm 0.0mm, clip, width=0.85\textwidth]{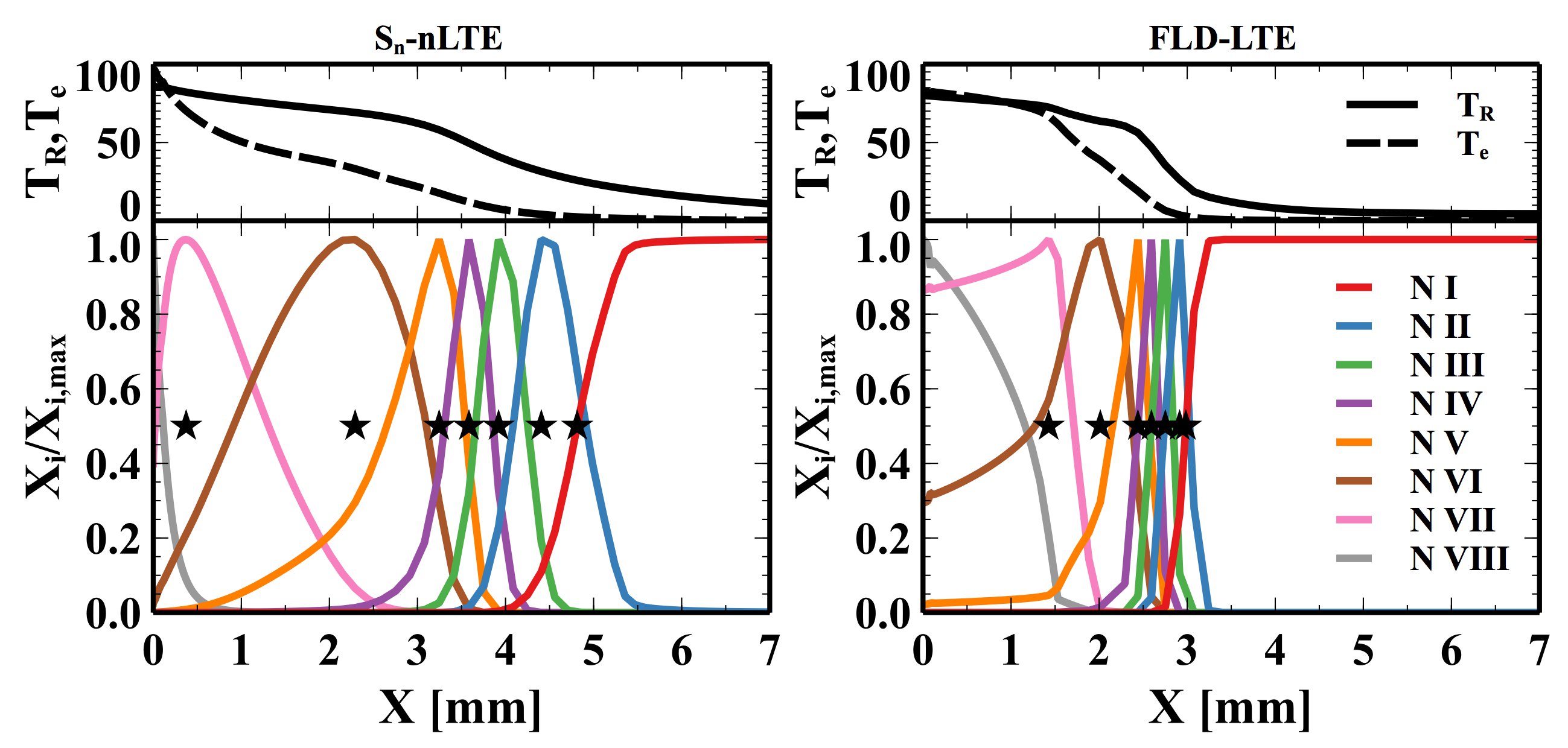}
\caption{ Normalized ionization state fractions for {\it left panel: } the nominal run and {\it right panel: } FLD-LTE at $t$=2 ns. The smaller upper panels show the electron temperature and radiation temperature. The legend gives the line color for each ionization state. The black stars show the positions where each ionization state is maximum.}
\label{fig:atomicfracs}
\end{center}
\end{figure*}

\begin{figure*}
\begin{center}
\includegraphics[trim=0.0mm 0.0mm 0.0mm 0.0mm, clip, width=0.85\textwidth]{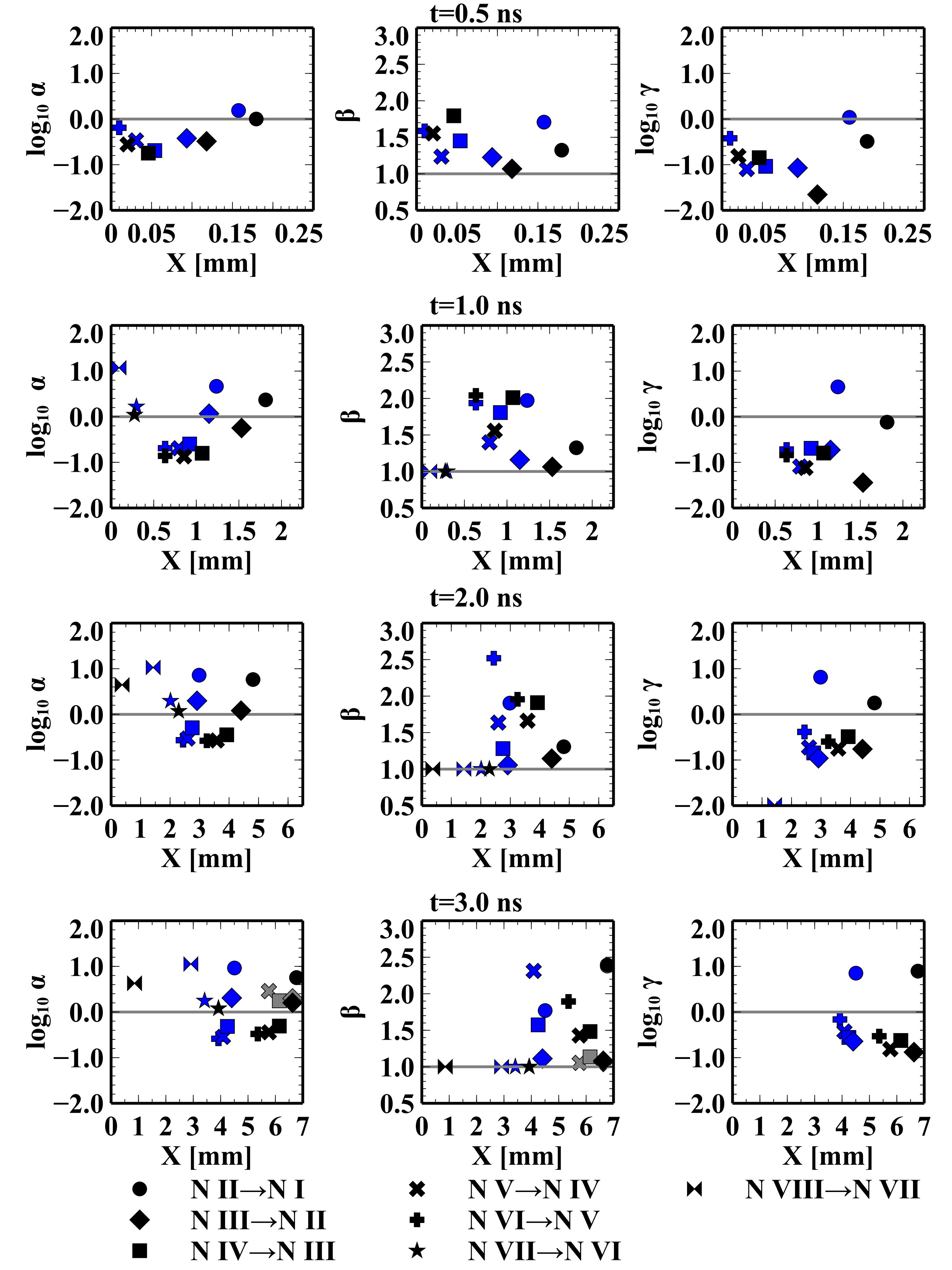}
\caption{ $\alpha$, $\beta$, and $\gamma$ values for the nominal model at four times.{\it top panel: } at $t$=0.5 ns, {\it upper middle panel: } $t$=1.0 ns, {\it bottom middle panel: } $t$=2.0 ns, and {\it bottom panel: } $t$=3.0 ns. For each time the left panel shows the $\alpha$ value, the middle panel shows $\beta$, and the right panel shows the $\gamma$ value. The black symbols represent results from S$_{n}$-nLTE, the blue symbols represent FLD-LTE, and the gray symbols in the bottom right panel show S$_{n}$-nLTE results including dielectronic recombination. Gray lines in each panel highlight $\alpha$=1, $\beta$=1, and $\gamma$=1 and are there to guide the eye. Note that $\alpha$ and $\gamma$ are plotted logarithmically while $\beta$ is linear.}
\label{fig:fidAlphaBeta}
\end{center}
\end{figure*}

The ultimate goal of this work and the companion experimental campaign \citep{LeFevre2018} is the creation and study of a photoionization front. In a previous work, Gray \etal \cite{Gray2018}, we presented results using the multi-material radiation hydrodynamics code \crash.\citep{vanderHolst2011} There we showed that photoionization fronts are possible over a wide range of peak radiation source temperatures and nitrogen gas pressures. Here we perform similar analysis as done in \cite{Gray2018} in order to determine if and when a photoionization front is formed when more sophisticated radiation transport and atomic kinetics methods are employed. The simulations are described in Table~\ref{tab:runsummary}.

\subsection{Dimensionless Parameters for Photoionization Fronts}

As discussed in Drake \etal \cite{Drake2016} and Gray \etal \cite{Gray2018}, a photoionization front can be described by two dimensionless quantities initially defined, for a simple two-state model, as,
\begin{equation}
\alpha_{i} = R_{i+1,i} n_{T} / \Gamma_{i,i+1}
\end{equation}
\begin{equation}
\beta_{i} = \sigma_{i,i+1}/R_{i+1,i} + 1,
\end{equation}
where $R_{i+1,i}$ is the total recombination rate coefficient, defined in Eq.~\ref{eqn:recombination} below, n$_{T}$ is the total number density of the gas, and $\sigma_{i,i+1}$ and $\Gamma_{i,i+1}$ is the electron ionization rate coefficient and local photoionization rate between state $i$ and $i+1$ respectively. For multi electron atoms, there is a unique $\alpha$ and $\beta$ for any two adjacent energy levels. In the case of nitrogen, there are seven distinct $\alpha$ and $\beta$ values.

These definitions proved useful in simplifying the analytic expressions found in Drake \etal \cite{Drake2016} and Gray \etal \cite{Gray2018}, but do not fully capture the physics involved. The time rate of change for a given ionization state is given by
\begin{equation}
\label{eqn:dndt}
\frac{dn_{i}}{dt} = -\sigma_{i,i+1}n_{e}n_{i}-\Gamma_{i,i+1}n_{i}+R_{i+1,i}n_{e}n_{i+1},
\end{equation}
with variable definitions given above. It is apparent where the definitions of $\alpha$ and $\beta$ are derived from. However, we note that the original definitions of $\alpha$ and $\beta$ do not have the number density factors as shown in Eq.~\ref{eqn:dndt}. More physically relevant definitions are defined as 
\begin{equation}
\label{eqn:palpha}
\alpha_{i} = \frac{n_{i+1}}{n_{i}} \frac{n_{e}R_{i+1,i}}{\Gamma_{i,i+1}}
\end{equation}
\begin{equation}
\label{eqn:pbeta}
\beta_{i} = \frac{n_{i}}{n_{i+1}} \frac{\sigma_{i,i+1}}{R_{i+1,i}}+1.
\end{equation}
Finally, an additional dimensionless quantity can be constructed from $\alpha$ and $\beta$, defined as 
\begin{equation}
\label{eqn:pgamma}
\gamma_{i} = \alpha_{i}(\beta_{i}-1) = \frac{n_{e}\sigma_{i,i+1}}{\Gamma_{i,i+1}},
\end{equation}
which relates the electron impact ionization rate to the photoionization rate. We make use of these definitions in the analysis and discussion presented below.

A photoionization front exists when $\alpha<<1$, $\beta\sim$1 and $\gamma<<1$. That is, we strive for an environment where the photoionization rate is much larger than the total recombination rate and the nitrogen is progressively more ionized. In addition, we want an environment where the rate of electron impact ionization is smaller than the total recombination rate.  
In astrophysical systems, typical values might be $\alpha=\gamma=$10$^{-4}$ and $\beta$=1. In laboratory systems typical values are likely closer to $\alpha=\gamma=$10$^{-1}$-10$^{-2}$ and $\beta\sim$1.
As we will show below, many ionization states have $\alpha$ values near 0.1 and $\beta$ values between 1 and 3. However, as discussed above $\beta$ should be nearly identical to one. This highlights the importance of $\gamma$ as it compares the electron impact ionization rate to the photoionization rate. If $\gamma<1$, then ionization by photons dominates while electron impact ionization dominates for $\gamma>1$. For nearly all of the cases explored below, we find that when $\beta\sim1-3$, $\gamma$ is $\sim$10$^{-1}$ which suggests that photons dominate the atomic kinetics of the gas.

We follow the procedure discussed in Gray \etal \cite{Gray2018} to compute $\alpha$ and $\beta$, which we summarize here. The total recombination rate from ionization state $i$ to state $j$ is given by
\begin{equation}
\label{eqn:recombination}
R_{ij} =  K_{ij} + D_{ij} + T_{ij}n_e,
\end{equation}
where K$_{ij}$ is the two body radiative recombination coefficient, D$_{ij}$ is the two-body dielectronic recombination coefficient, T$_{ij}$ is the three-body recombination coefficient, and n$_e$ is the number density of electrons. Radiative recombination coefficients are computed using the rates presented in Badnell \etal \cite{Badnell2006}, dielectronic recombination rate coefficients are computed using results from a series of papers that generated the dielectronic recombination coefficients for use in the study of astrophysical and laboratory plasmas. \citep{Badnell2003}\citep[\eg][]{Colgan2003,Colgan2004,Zatsarinny2004b,Altun2004,Badnell2006,Bautista2007} \cite{Nikolic2013} showed that for high-density plasmas, n$_{e}>$10$^{10}$cm$^{-3}$, dielectronic recombination rates are highly suppressed. As all the models presented here are above this threshold, we compute Eq.~\ref{eqn:recombination} with and without the dielectronic contribution and comment below on its importance. Three-body recombination rate coefficients are taken from Drake \etal  \cite{Drake2016} and Lotz\etal\cite{Lotz1967} Drake \etal \cite{Drake2016} showed that for laboratory photoionization front experiments that three-body recombination is not a dominant source of recombinations. We therefore choose to slightly overstate the three-body contribution by replacing the electron number density with $Zn_{T}$, where $Z=7$ for nitrogen and $n_{T}$ is the initial nitrogen number density within the gas cell. Electron impact ionization coefficients are computed using fits presented in Voronov \etal\cite{Voronov1997}

\subsection{Photoionization Rate}

Before presenting results for $\alpha$ and $\beta$, it is prudent to discuss how the photoionization rate should be computed. In our previous work, \citep{Gray2018}, the photoionization rate was computed as
\begin{equation}
\Gamma=\overline{\sigma}_{\gamma}F_{\gamma},
\end{equation}
where $\overline{\sigma}_{\gamma}$ is the spectrally averaged photoionization cross section and F$_\gamma$ is the photon flux. The photon flux was computed using a FLD approximation using the `square root limiter'.\citep{vanderHolst2011} This was appropriate as the simulations presented in Gray \etal\cite{Gray2018} employed the same FLD approximation. However, the simulations presented here make use of a multi-angle long characteristic model and instead we seek a model that is more physically accurate.

The local photoionization rate, $\Gamma(z)$, is given by the following integral: 
\begin{align} 
\label{eqn:PI1}
\Gamma (z) =  \int_0^\infty \frac{E_{R,\nu}(z) c}{h \nu}  \sigma_\nu \D \nu,  
\end{align}
in which $\sigma_\nu$ is the photoionization cross section of interest at frequency $\nu$, the spectral radiation energy density at $\nu$ and $z$ is $E_{R,\nu}(z)$, $h$ is the Planck constant, and $c$ is the speed of light. In the limit that the radiation is beamlike, the quantity $E_{R,\nu} c = F_R$, is the radiation flux. The radiation energy density develops in consequence of photon emission followed by absorption en route to any given location of interest. 

Our goal in the present section is to find a way to obtain a reasonable estimate for the photoionization rate as defined in Eq.~\ref{eqn:PI1}. We will use this for a comparison of the rates of atomic processes, to find ratios that establish the regime of a given physical system. We will be interested in variations of these ratios across several orders of magnitude. 

We begin with the total radiation energy density, $E_R$. The total radiation energy density is 
\begin{equation} 
E_{R}(z) = \int_{0}^{\infty} E_{R,\nu}(z) \D \nu,
\label{eqn:PI7}
\end{equation}
Typical code output, including that from \helios\, provides $T_R$, as shown in the figures above, inferred from $E_R$ on the assumption of thermal equilibrium. However, the local spectrum often is not Planckian. In the present case, we can find a more accurate approximation of the spectrum of the radiation by considering some fundamental aspects of radiation transport. 

Given a distribution of plasma parameters, including local electron temperature $T_e(z)$, and a source temperature $T_s$, the radiation energy density produced at $z$ by the source is 
\begin{equation}
E_{R,\nu,s}(z) = 2 \pi \int_0^1 I_{\nu,\mu}(z)  \D \mu ,  
\label{eqn:PI2}
\end{equation} 
in which $\mu = \cos \theta$ and, for this case of forward going radiation only, the integral is from 0 to 1. Here the spectral intensity reaching the plane at $z$ from the source, and propagating with angle variable $\mu = \cos \theta$, is 
\begin{equation}
\resizebox{0.8\columnwidth}{!}
{
 $I_{\nu,\mu}(z)  =   B_\nu(T_s) \exp \left[ -\int_0^z \frac{\kappa_{\nu,z'}}{\mu} \D z' \right]$ ,
}
\label{eqn:PI3}
\end{equation} 
in which $B_\nu$ is the thermal spectral intensity and $\kappa_{\nu,z}$ is the opacity at frequency $\nu$ and location $z$, based on the plasma parameters.

We anticipate that the photon density resulting from thermal emission throughout the plasma is small, as the plasma is either cold or optically thin. But it can be calculated as the sum of downstream and upstream contributions. This is  
\begin{equation}
\resizebox{0.85\columnwidth}{!}
{
$I_{\nu,\mu, th}(z)  = \int_0^z  B_\nu(T_e(z_o)) 
\exp \left[ -\int_{z_o}^z \frac{\kappa_{\nu,z'}}{\mu} dz'  \right] \kappa_{\nu,{z_o}}  \frac{\D z_o}{\mu}$
}
\label{eqn:PI4}
\end{equation} 
for $\mu > 0$, plus 
\begin{equation}
\resizebox{0.85\columnwidth}{!}
{
$I_{\nu,\mu, th}(z)  = \int_z^\infty  B_\nu(T_e(z_o)) 
\exp \left[ \int_{z}^{z_o} \frac{\kappa_{\nu,z'}}{\mu} dz'  \right] \kappa_{\nu,{z_o}}  \frac{\D z_o}{\mu}$
}
\label{eqn:PI5}
\end{equation} 
for radiation from upstream having $\mu < 0$. From these, the radiation energy density from thermal emission is 
\begin{equation}
E_{R,\nu,th}(z) = 2 \pi \int_{-1}^1 I_{\nu,\mu,th}(z)  \D \mu .  
\label{eqn:PI6}
\end{equation} 
In general, we now have $E_{R,\nu}(z) = E_{R,\nu,s}(z) +E_{R,\nu,th}(z)$, but the thermal emission does indeed turn out to be negligible, as shown by the light gray curve in Fig.~\ref{fig:raytrace}. 

We want to use Eqns.~\ref{eqn:PI7} to~\ref{eqn:PI6} to evaluate the radiation energy densities, in order to understand the origin of the results from \helios. To do so, one needs plasma parameters and opacities.  We obtain the plasma parameters from \SN\ at $t=$1 ns, where we have the spatial profiles for the radiation temperature, electron temperature, and mass density. However, we must find some approximation for the opacities, as we do not have the many, frequency-dependent values calculated by the inline model in the code. 

To obtain opacities for the calculation, we interpolate a table provided by \propaceos\ for LTE conditions. This table is defined on 50 points in number density between { 4$\times$10$^{18}$ and 4$\times$10$^{21}$ cm$^{-3}$} and 150 points in electron temperature ranging between 0.01 and 120 eV. Fifty radiation groups are defined between 0.1 and 120 eV.  Finally, $\mu$ is defined on a grid of 21 points ranging between 0 and 1. We anticipate that the results of calculations using this opacity table will diverge strongly from those of the nLTE runs when $T_R$ begins to significantly exceed $T_e$, producing opacities that diverge strongly from the LTE values. 

Figure~\ref{fig:raytrace} shows the results of this procedure. The left and middle panel shows the radiation energy density and radiation temperature respectively, from the calculation just described and from the \SN\ and S$_{n}$-LTE models from \helios. For the \helios\ models, radiation energy density is computed as $E_{R}=a$T$_{\rm R}^4$, where $a$ is the radiation constant. We find that our ray tracing procedure is able to reproduce the \helios\ results to within a factor of two over the region of interest, which extends from the left boundary to the dashed black line. This line shows the ionization fraction of neutral nitrogen. Everything to the right of this line is completely neutral while everything to the left is ionized. We also find that the self-emission of the gas, shown by the gray curve, is small compared to the absorbed component. We also find the expected divergence of the approximate model from the S$_{n}$-LTE model as one enters the region where $T_R$ begins to significantly exceed $T_e$.

The point of this exercise is that is demonstrates that the radiation energy density at a given location in the \helios\ \SN\ models is produced by the gradual attenuation of the emission from the source. This confirms what one would expect from simple calculations. 

The photoionization rate was defined above in Eq.~\ref{eqn:PI1}.  
Since the ray trace procedure shows that the radiation present at any given location is produced by transmission from the source, we can infer that the spectral temperature of this radiation will approximately equal that of the source.  We correspondingly approximate $cE_{R,\nu}$ as 
\begin{equation}
\label{eqn:PhotoIni2}
c E_{R,\nu} = 4 \pi \left(\frac{T_{R,l}}{T_{R,s}}\right)^4 B_{\nu}(T_{R,s}),
\end{equation}
where T$_{R,s}$ is the radiation temperature at the source of the emission, T$_{R,l}$ is the local radiation temperature from the \helios\ output, and B$_{\nu}$ is the Planckian thermal intensity. This has the effect of defining the spectral shape by the source radiation temperature and scaling the energy density by the local radiation temperature. Then, to calculate the local photoionization rate Eq.~\ref{eqn:PI1} is computed numerically with Eq.~\ref{eqn:PhotoIni2} and photoionization cross sections taken from Verner \etal\cite{Verner1995,Verner1996}

For comparison with our previous results, we evaluated the radiation flux as a function of frequency, computed as
\begin{equation}
F_{R,\nu}(z) = 2 \pi \int_{-1}^1 I_{\nu,\mu}(z)\mu  \D \mu,
\label{eqn:PI8}
\end{equation}
and the total radiation flux from 
\begin{equation}
F_{R}(z) = \int_{0}^{\infty} F_{R,\nu}(z) \D \nu.
\label{eqn:PI9}
\end{equation}
The right panel of Figure~\ref{fig:raytrace} shows the radiation photon flux as computed by Eq.~\ref{eqn:PI8} and Eq.~\ref{eqn:PI9} using the same parameters as for the prior calculations. The photon flux for the \helios\ models is estimated as $F_{\gamma}=2.36\times$10$^{23}$ T$_{\rm R}^3$ cm$^{-2}$s$^{-1}$, as reported by Drake \etal\cite{Drake2016} The result from our ray trace procedure is comparable to the results from \SN, until the LTE opacity becomes inaccurate as discussed above. The purple line shows, for comparison, the flux computed using a flux-limited diffusion approximation, shown as the purple line. This calculation used a `square-root' limiter \citep{vanderHolst2011} with Rosseland mean opacities from \SN. As shown in the figure, the FLD flux drastically under predicts the radiation flux in the system. This is the main origin of the differences between the results shown below and those of Gray \etal\cite{Gray2018}

\subsection{Location of the Photoionization Front}
There is no clear ideal method of determining where the photoionization front is located. 
The method described in Gray \etal \cite{Gray2018} was to find the location where a given ionization state has fallen to 50\% of its maximum value. 
Here we use a slightly different method. 
The top left panel of Figure~\ref{fig:alphabetapos} shows the ionization state fractions where the black star represents the maximum value for the \NVI\ ionization state. 
The top right panel shows $\alpha$, the bottom left shows $\beta$, and the bottom right shows $\gamma$ for each ionization state. 

For a given ionization state there are three distinct regions are seen in the profiles of $\alpha$, a region of high values near the source, a region where $\alpha$ reaches a minimum, and a region where $\alpha$ rises toward the end of the gas cell.
The first region is due to the high recombination rate as the ratio of ionization state densities ($n_{i+1}/n_{i}$) is small and the very high electron densities.
The second region is where the ionization state ratio is dominated by n$_{i}$ that drives down $\alpha$. 
Once the ratio equalizes and n$_{i+1}$ dominates $\alpha$ rises and begins to plateau.
Farther down the gas cell the ionization state ratio again becomes unimportant and is driven by the relatively low recombination rates at lower electron temperatures. 

Due to the inverse dependence on n$_{i}$ and n$_{i+1}$, $\beta$ shows a slight inverse profiles as found in $\alpha$. 
That is, where a minimum is found in $\alpha$, there is slight maximum found in $\beta$. 
However, we find that the range over which $\beta$ varies is much smaller than that found in $\alpha$.
In fact, for most ionization states $\beta$ does not exceed $\sim$3.
Only \NV\ shows a dramatic rise in $\beta$ that is attributed to the wide ionization state distribution. 

$\gamma_{i}$ is shown in the bottom right panel of Figure~\ref{fig:alphabetapos}. 
Since this value does not depend on the ratio of ionization states, it does not contain the minimum or maximum found in $\alpha$ or $\beta$.
For each ionization state, $\gamma$ is very large toward the source and is due to both the increase in the electron number density and the strong temperature dependence on the electron impact ionization rate coefficients. 
However, even when the ionization fractions are at their peak $\gamma$ is less than one and shows that each ionization state is photoionized at some point within the gas cell.
For example, for \NV\ at is peak, we find that photoionization accounts for two out of every three ionization events.

Finally, the color-coded bars above each ionization state in Figure~\ref{fig:alphabetapos} represent the mean free path ($\ell_{i} = 1/n_{i}\sigma_{i}$) at the peak of each ionization state. 
As discussed in Drake \etal \cite{Drake2016}, a successful photoionization experiment will consist of a system with tens to hundreds of mean free paths. 
Except for \NVII, all ionization state show mean free paths that are small compared to the system size. 
The computed mean free paths are on the order of $\ell_{i}\sim$0.03-0.3 mm, equaling 30 to 300 mean free paths and satisfy above condition on the system size.

Figure~\ref{fig:atomicfracs} shows the results of using the maximum value at $t$=2 ns for both the nominal model (\SN) and FLD-LTE. The difference between LTE and nLTE in terms of the distribution of ionization states is quite striking. The nLTE model produces a much broader distribution for each ionization state compared to the LTE model. In addition, the nitrogen gas has begun to ionize much further into the gas cell, with \NII\ appearing at 5 mm down the cell. FLD-LTE, on the other hand, does not have \NII\ appearing until 3 mm within the cell. Shown above each panel is the electron and radiation temperature for each simulation. At each of these points all the required physical data needed to compute $\alpha$, $\beta$, and $\gamma$ is either defined or computed using the above definitions. 

If a given ionization state is unpopulated  $\alpha$ and $\beta$ are set to large values. This can occur for high ionization potential states early in time or for low ionization potential states late in time if the gas is ionized beyond a given ionization state. 

\subsection{Photoionization Front Results}
Figure~\ref{fig:fidAlphaBeta} shows the $\alpha$, $\beta$, and $\gamma$ values for S$_{n}$-nLTE and FLD-LTE as a function of position in the gas cell at four times. 
The location of each point is chosen as described above, that is, we choose the location where each ionization state is at a maximum.
At early times, $t=$0.5 ns and shown in the top panel, both \SN\ or FLD-LTE have $\alpha$, $\beta$, and $\gamma$ values that are consistent with a photoionization front. 
Only \NI\ produces $\alpha$ values that are near one. 
For all populated ionization states, $\beta$ is near 1.
$\gamma$ shows values near 0.1, which suggests that most ionization events are due to photons.
In addition, \SN\ and FLD-LTE produce $\alpha$ and $\beta$ values that are roughly consistent with each other.

This remains largely true at $t=$1.0 ns which is shows in the upper middle panel. 
$\alpha$ values between 10$^{-1}$ and 1 are found for all ionization states except for \NI\ and \NII. 
In all cases, $\beta$ remains near between one and two. 
Except for \NI\ $\gamma$ remains less than one. 
Here we start to see a difference between \SN\ and FLD-LTE with higher ionization states appearing deeper into the gas cell for \SN. 
However, $\alpha$, $\beta$, and $\gamma$ values are still roughly consistent between the two models.

The bottom middle panel of Figure~\ref{fig:fidAlphaBeta} shows the $\alpha$ and $\beta$ values at $t=$2.0 ns. 
Ionization states between \NII\ and \NVIII\ have $\alpha<1$, $\beta\sim$1-1.5, and $\gamma<$1 and is indicative of a photoionization front. 
Slightly lower values of $\alpha$ are found for \SN\ compared to FLD-LTE while $\beta$ and $\gamma$ values remain consistent.
The separation between \SN\ values and FLD-LTE values continues to widen.  

Values of $\alpha$, $\beta$, and $\gamma$ at $t=$3.0 ns are shown in the bottom panel of Figure~\ref{fig:fidAlphaBeta}. 
As in the previous panels, we find $\alpha$, $\beta$, and $\gamma$ values that are consistent with a photoionization front for moderately ionized species of nitrogen. 
In most cases, $\alpha$ lies between 10$^{-1}$ and one, $\beta$ values near one, and $\gamma$ values near 10$^{-1}$. 
The gray symbols in Figure~\ref{fig:fidAlphaBeta} show the results when dielectronic recombination is included. 
For most ionization states, this has little to no effect on either $\alpha$, $\beta$, or $\gamma$. 
The inclusion of dielectronic rates tends to slightly increase the total recombination rate that slightly raises $\alpha$ and slightly lowers $\beta$. 
However, this effect is minor and does not drastically change the interpretation of $\alpha$ or $\beta$.

Taken together, Figs~\ref{fig:atomicfracs} and \ref{fig:fidAlphaBeta} show that the dominant ionization state, \NVI, is photoionized between $t=$1-3 nanoseconds with $\beta$$\sim$1, $\alpha\sim$10$^{-1}$, and $\gamma\sim$10$^{-1}$.
This is true even at late times, $t=$3 ns, even though the flux has begun to drop due to optical effects through the gas. 
In the following two subsections, we study the importance of the drive temperature and the nitrogen gas pressure on the creation of a photoionization front.

\subsection{Effect of Drive Temperature}
\label{sec:DriveTemp}

\begin{figure*}
\begin{center}
\includegraphics[trim=0.0mm 0.0mm 0.0mm 0.0mm, clip, width=0.85\textwidth]{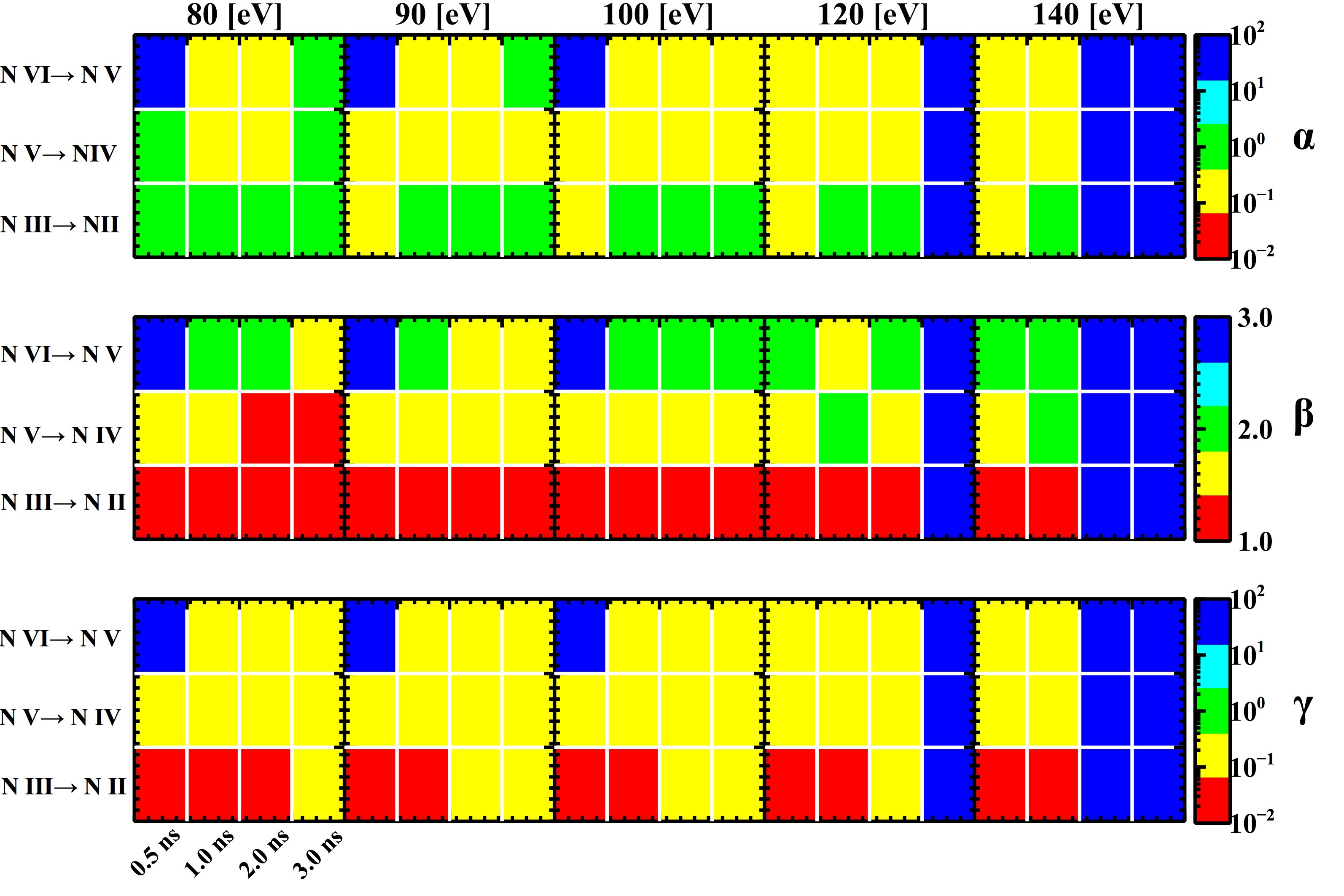}
\caption{Results from varying the peak radiation temperature at the boundary. $\alpha$ is shown in the top panel, $\beta$ is shown in the middle panel, and $\gamma$ is shown in the bottom panel. Each simulation is given by a four by three panel with time given along the $x$-axis and atomic transition given along the $y$-axis. Each column gives the results for a given simulation with the peak temperature given as the column label. $\alpha$ and $\beta$ values are color coded by the given color bars. Note that the logarithm of $\alpha$ and $\gamma$ is plotted while $\beta$ is linear. }
\label{fig:compDriveTemp}
\end{center}
\end{figure*}

Four simulations were performed that varied the peak radiation temperature boundary condition between 80 eV and 140 eV. The nitrogen pressure is kept constant at ten atmospheres. We concentrate on three atomic transitions, \NIII$\rightarrow$\NII, \NV$\rightarrow$\NIV, and \NVI$\rightarrow$\NV. \NIII$\rightarrow$\NII\ and \NV$\rightarrow$\NIV\ are chosen to represent transitions for the 2p subshell and 2s subshell. \NVI$\rightarrow$\NV\ is chosen since \NVI\ is the dominant ionization state reached in the nominal simulation. Four times were chosen: $t=$0.5 ns as the source temperature ramps up, $t=$1.0 ns where the source reaches peak, and $t=$2 and $t=$3 ns where the source is emitting at its full potential. 

Figure~\ref{fig:compDriveTemp} shows the $\alpha$, $\beta$ and $\gamma$ results for these simulations. 
A four-by-three panel describes each simulation with time running along the x-axis and atomic transition running along the y-axis. 
The value of $\alpha$, $\beta$, and $\gamma$ is given by the color bar. 
Each column represents a different peak radiation temperature boundary that is given in the title. 

The \NIII$\rightarrow$\NII\ transition has favorable $\beta$ and $\gamma$ values and marginal $\alpha$ values for all peak radiation temperatures at $t=$0.5 ns. 
At later times electron impaction ionization begins to become more important, producing values of $\alpha$ near one. 
This is due to the fact that \NII\ has a very high electron impact ionization rate even at very low electron temperatures. 
For very high peak temperatures the front is able to ionize the entire gas cell beyond \NIII\ that accounts for the large values of $\alpha$, $\beta$, and $\gamma$ at late times. 

\NV$\rightarrow$\NIV, on the other hand, produces $\alpha$ values consistent with a photoionization front over the range of peak radiation temperatures. 
$\beta$ values alway remain near one. 
As seen in the \NIII$\rightarrow$\NII\ case, for high radiation temperatures no front is produced. 
This is due to the gas cell being completely ionized beyond the these ionization states.
$\gamma$ is also consistent with a photoionization front, with values near 10$^{-1}$.

\NVI$\rightarrow$\NV\ is very similar to \NV$\rightarrow$\NIV\ but with slightly more favorable $\alpha$ values. 
In fact, over a wide range of boundary radiation temperatures and times, we find that $\alpha\sim$10$^{-1}$. 
$\beta$ values are slightly higher than in the \NV$\rightarrow$\NIV, with values closer to two.
However, $\gamma$ suggests that most ionization events are driven by photon, with values close to 10$^{-1}$.

Across Fig~\ref{fig:compDriveTemp} one sees large variations in $\alpha$, $\beta$, and $\gamma$. Under some circumstances, the calues are large (shown in blue), indicating that photoionization is negligible. At early times, for \TR\ up to 100 eV, the \NVI\ has yet to be excited. At late times and high radiation drive temperatures, this occurs when the entire gas cell has become ionized above \NVI.

One important experimental consideration presented by Figure~\ref{fig:compDriveTemp} is that whether or not a photoionization front is produced is not strongly dependent on the peak radiation temperature at the entrance to the nitrogen gas cell. In fact, for nearly every transition there is a time where the conditions are favorable for a photoionization front. This is particularly important for \NVI\ which is the dominant species formed.

\subsection{Effect of Nitrogen Pressure}
\label{sec:NPressure}
\begin{figure*}
\begin{center}
\includegraphics[trim=0.0mm 0.0mm 0.0mm 0.0mm, clip, width=0.85\textwidth]{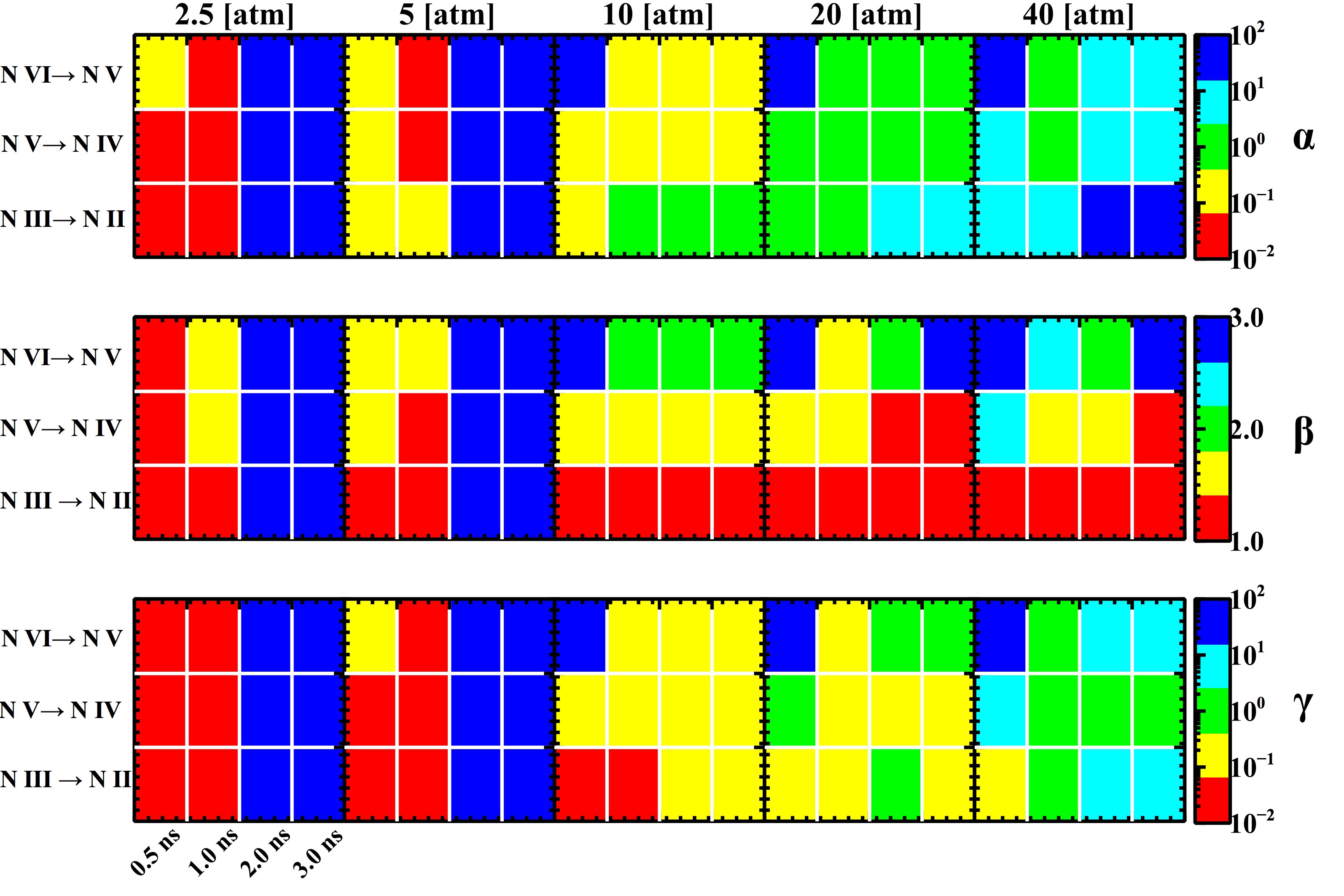}
\caption{Results from varying the initial nitrogen gas pressure. $\alpha$ is shown in the top panel, $\beta$ is shown in the middle panel, and $\gamma$ is shown in the bottom panel. Each simulation is given by a four by three panel with time given along the $x$-axis and atomic transition given along the $y$-axis. Each column gives the results for a given simulation with the initial gas pressure given as the column label. $\alpha$, $\beta$, and $\gamma$ values are color coded by the given color bars. Note that the logarithm of $\alpha$ and $\gamma$ is plotted while $\beta$ is linear.}
\label{fig:compNPressure}
\end{center}
\end{figure*}

An additional four simulations are run in order to study the effect of the nitrogen pressure on the formation of a photoionization front. As described in Table~\ref{tab:runsummary}, the nitrogen pressure is varied between 2.5 and 40 atmospheres with the peak radiation temperature at the boundary fixed at 100 eV. 

Figure~\ref{fig:compNPressure} shows the results of these models. As in figure~\ref{fig:compDriveTemp} each 4 by 3 panel shows the $\alpha$, $\beta$, and $\gamma$ values for several ionization state transitions at different times. Ionization state transition is given along the $y$-axis and time along the $x$-axis. The nitrogen pressure is given in the title above each column. 

At low pressures, between 2.5 and 5 atmospheres, the front quickly moves through the gas cell. The front velocity is given by $v_{f}=F_{\gamma}/n$ where F$_{\gamma}$ is the photon flux and $n$ is the particle number density. For a given flux the front moves faster for lower nitrogen pressures and quickly ionizes the gas. The transitional length of the photoionization front is a few mean free paths for the initial photoionization, where the mean free path is given by $\lambda_{mfp} = 1/n\sigma$ where $n$ is the particle number density and $\sigma$ is the photoionization cross section. As discussed above, an ideal experiment would then require the ratio of the experiment size to the mean free path to be large. As shown in figure~\ref{fig:compNPressure} it is likely that 2.5 and 5 atmospheres provides too few mean free paths for a photoionization experiment on a facility such as Omega where the spatial extent of the system can only be a few mm. 

Nitrogen pressures of 10 and 20 atmospheres, on the other hand, provide a more ideal environment for the creation of a photoionization front. For \NV$\rightarrow$\NIV\ $\alpha<1$ at all time for 10 atmospheres and before 1 ns in the 20 atmosphere case. Lower ionization states aren't as favored as recombinations start to dominate. For higher pressures the larger ionization states are favored for an photoionization front. At very high nitrogen pressures it becomes very difficult to form a photoionization front. This is due to three-body recombination rates and their strong dependence on the electron number density.

Similar to the radiation drive temperature comparison, large values of $\alpha$, $\beta$, and $\gamma$ are found in some cases. 
At low pressures, 2.5$<$P$<$5 atm, this occurs at late times because the ionization front has quickly swept through the gas cell and ionized beyond those transitions. 
At high pressures, 10$<$P$<$40 atm, \NV$\rightarrow$\NIV\ shows these large values early in time because these ionization states are not yet excited.
In addition, \NIII$\rightarrow$\NII\ also shows large values of $\alpha$ at high pressure, which is due to large recombination rates and low photoionization rates.

When compared to varying the peak radiation drive temperature the nitrogen pressure is more important in the formation of a photoionization front. For low nitrogen pressures the front moves very quickly through the gas cell as there are too few mean free paths to resolve the photoionization front. At very high pressures three-body recombinations dominate and a electron heat front is generated. For the proposed photoionization front experiment, a nitrogen pressure between 5 and 20 atmospheres is feasible while a pressure between 5 and 10 atmospheres is ideal.

\section{Summary and Conclusion}

We have presented a suite of one-dimensional \helios\ simulations in order to study the formation of a photoionization front in a nitrogen gas cell. 
This work builds upon the theoretical analysis done by Drake \etal \cite{Drake2016} and the two dimensional numerical simulations of Gray \etal\cite{Gray2018}
By using \helios\ we are able to study the effect of changing the complexity of the physics in the formation of a photoionization front, which we were unable to do in the previous two-dimensional models.
Specifically, we are able to study the effect of changing the radiation transport method between a flux limited diffusion model and a more complex multi-angle long characteristics (S$_{n}$) model.
The atomic kinetics model is also varied between a local thermodynamics equilibrium model (LTE) and a collisional-radiative model (nLTE). 

The mass density, radiation temperature, electron temperature, and nitrogen ionization state were compared for different radiative transfer and atomic kinetics models. We find that there is little difference between S$_{n}$ and FLD radiative transfer models when run with LTE conditions. The atomic kinetics model, however, has a dramatic effect on the ionization state of the nitrogen gas. In particular, the radiation and electron temperatures are higher deeper within the nitrogen gas for \SN\ compared to either FLD-LTE or S$_{n}$-LTE models. This creates higher average ionization states within this warmer gas.

To determine whether or not a photoionization front is formed in our simulations we compute three dimensionless values: $\alpha$, which relates the total recombination rate to the photoionization rate, $\beta$, which relates the electron impact ionization rate to the total recombination rate, and $\gamma$ which relates the electron impact ionization rate to the photoionization rate. We find that for our nominal model with ten atmospheres of nitrogen and a radiation source temperature of 100 eV that a photoionization front is formed for several ionization states of nitrogen up to 3 ns after the source is turned on.

We also present a suite of simulations that vary the source radiation temperature and nitrogen pressure to determine if a photoionization front forms. One set of simulations fixed the nitrogen pressure at ten atmospheres and varied the source radiation temperature while the other fixed the temperature at 100 eV and varied the nitrogen pressure. We find that the source radiation temperature does not strongly impact the formation of a photoionization front, with photoionization fronts forming for peak radiation temperatures between 80 and 100 eV. The nitrogen pressure plays a more dominant role with nitrogen pressures between 5 and 20 atmospheres being ideal.

We find that the results presented here match those presented in Gray \etal\cite{Gray2018}
In the simulation presented there, the photoionization front experiment was modeled in two dimensions and employed flux limited diffusion and local thermodynamic equilibrium for its radiative transfer and atomic kinetics respectively. 
In this previous work, we found that a photoionization front was expected for radiation temperature above 90 eV and nitrogen pressure between 5 and 20 atm.
Although the specific values of $\alpha$ and $\beta$ differ between the two studies, we find similar results here.

The simulations presented here provide important insight into the creation of photoionization fronts in nitrogen gas and future laboratory experiments. We conclude that with the inclusion of more detailed physics that photoionization fronts likely to form in the proposed laboratory experiment. Future work and simulations will be able to provide a more favorable parameter space where photoionization fronts can be formed.

\acknowledgments
R.P.D and W.J.G were supported by the U.S. Department of Energy, through the NNSA-DS and SC-OFES Joint Program in High-Energy-Density Laboratory Plasmas, grant number DE-NA0002956, and by the Lawrence Livermore National Laboratory under subcontract B614207. Helpful comments by the referee are also gratefully acknowledged.


\bibliographystyle{elsarticle-num}
\bibliography{ms.bib}

\end{document}